\documentclass[a4paper,left=2cm.right=2cm,top=2cm,bottom=2cm,11pt]{article}
\usepackage{amsmath}

\begin{document}
\title{Cosmological Model with Nonminimal Derivative Coupling of Scalar Fields in Five Dimensions}
\author{Agus Suroso and Freddy P. Zen\\
Theoretical Physics Laboratory, THEPI Division, and \\
Indonesia Center for Theoretical and Mathematical Physics (ICTMP) \\
Faculty of Mathematics and Natural Sciences, Institut Teknologi Bandung\\
Jl. Ganesha 10 Bandung 40132, Indonesia.\\
agussuroso@fi.itb.ac.id, fpzen@fi.itb.ac.id}

\maketitle

\begin{abstract}
We study a nonminimal derivative coupling (NMDC) of scalar field, where the scalar field is coupled to curvature tensor in  the five dimensional universal extra dimension model. We apply the Einstein equation  and find its solution. First, we consider a special case of pure free scalar field without NMDC and we find that for static extradimension, the solution is equivalent to the standard cosmology with stiff matter. For a general case of pure free scalar  field with NMDC, we find that the de Sitter solution is the solution of our model. For this solution, the scalar field evolves linearly in time. In the limit of small Hubble parameter, the general case give us the same solution as in the pure free scalar field. 
Finally, we perform a dynamical analysis to determine the stability of our model. We find that the extradimension, if it exist, can not be static and always shrinks with the expansion of four dimensional spacetime. 
\end{abstract}

\section{Introduction}
Extra dimensions have a special role in physics. 
At first, its existence was proposed to solve some theoretical problem
concerning unification of all physical interactions. Kaluza and Klein proposed one dimensional  extra dimension when developed a theory which  unified  gravity and electromagnetism \cite{Overduin:1998pn}. The string theory, as a good candidate of unification theory, requires 26 (for bosonic string) and 10 dimensions (for superstring) of spacetime in order to be consistent. The idea of \textit{brane} which exists in string theory then adopted by physicist to propose the braneworld gravity. The five dimensional braneworld gravity suggests that our four dimensional spacetime in constant time can be viewed as a 3D subspace called \emph{3-brane} embedded in five dimensional space called \emph{bulk}. The matter field is constrained on the brane and the gravitational field can move freely in the bulk  \cite{maartens}. The braneworld model has been studied by Lisa Randall and Raman Sundrum to explain the hierarchy problem between Planck scale and electrowake scale \cite{RSI,RSII}.
Gravity on the brane is described by the modified Einstein equation developed by Shiromizu, Maeda, and Sasaki (SMS) \cite{SMS}. This equation derived from the Einstein equation in the bulk using the Gauss-Codacci equation and Israel's junction condition of the brane. Different from the standard Einstein equation, this equation contains two additional terms related to quadratic term of energy-momentum tensor and the effects of the bulk geometry on the brane. Cosmological application of this equation has been studied extensively, for a review see \cite{Brax:2003fv,Langlois:2002bb}. Modification of the SMS equation with Lorentz invariance violation and its applications in cosmology has been studied in \cite{arianto1,arianto,arianto2,zen}. 

Another kind of extradimensional theory of gravity is the universal extra dimension (UED). This model  proposes that the matter field is not confined on the brane but can move freely in the bulk \cite{UED}. 
This model does not solve the hierarchy problem but provide a very good candidate for the cold dark matter (CDM) and solve the proton stability  problem  naturally \cite{Hsieh:2006qe,Appelquist:2001mj}. The cosmological aspect of this theory has been studied in \cite{Bringmann:2003sz,Bringmann:2004xt}.

In this paper, we study the cosmological aspect of the five dimensional extra dimension (5D UED) model, where the source of dynamics is given by a nonminimal derivative coupling (NMDC) of scalar field and the curvature tensor. 
The NMDC was introduced by Amendola in \cite{amendola}.  This model has been  studied in the context of inflation  and late time acceleration \cite{Capozziello:1999xt,capozziello2,granda}. The exact cosmological solution of NMDC in four dimensional spacetime has been studied in \cite{sushkov}. 

Here, we investigate how the NMDC terms influence the evolution of the four dimensional spacetime and the extra dimension. We  consider the NMDC as the quantity that cause the acceleration of the four dimensional spacetime and make the extradimension shrink to a small size. 
As the results, we see that a solution where the four dimensional spacetime expands with acceleration and the extra dimension shrinks in time 
can arise naturally in our model.

We organize this paper as follow. After this brief introduction, we setup the basic equations of the model. We give the geometrical quantities of the ansatz metric and setup the Einstein equation based on the Lagrangian of this model in the Section 2. Then, we solve the Einstein equation for a special cases of pure free scalar field and a general case of pure free scalar with pure NMDC. Next, we perform dynamical analysis to determine stability of our model in Section 4. Finally, we give the conclusions in the last section.

\section{Model Setup}
Lets start from the action
\begin{align}
S=\int d^5x \sqrt{-g} \left(\frac{R}{2\kappa^2}  +\frac{1}{2}g_{AB} \partial^A \phi \partial^B \phi + \frac{\xi}{2} R g_{AB} \partial^A \phi \partial^B \phi + \frac{\eta}{2} R_{AB}  {\partial^A \phi \partial^B \phi} \right).
\end{align}
Here, the nonminimal derivative coupling is given in the two last terms of the action. 

Taking a variation of this action with respect to the metric, 
we obtain the  Einstein equations  in five dimensional spaces as
\begin{equation}
\label{eq:einstein}
G_{AB} = \kappa^2 T^{(\phi)}_{AB},
\end{equation}
where 
\begin{align}
T^{(\phi)}_{AB} &= \nabla_A \phi \nabla_B \phi 
- \frac{1}{2} g_{AB} \nabla_C \phi \nabla^C \phi \nonumber\\
+ \xi &\left[G_{AB} \nabla^C\phi \nabla_C\phi 
+ R \nabla_A \phi \nabla_B\phi 
- \nabla_A \nabla_B \left(\nabla^C\phi \nabla_C\phi\right) \right.\nonumber\\
 \qquad&+\left. g_{AB} \nabla_C\nabla^C\left(\nabla_D\phi\nabla^D\phi\right)\right]\nonumber\\
+\eta&\left[R_{AC} \nabla_B \phi \nabla^C\phi 
+ R_{CB} \nabla^C\phi \nabla_A\phi 
-\frac{1}{2} g_{AB} R_{CD} \nabla^C\phi \nabla^D \phi\right.\nonumber\\
+ &\frac{1}{2} g_{AB} \nabla_C \nabla_D \left(\nabla^c\phi \nabla^D\phi\right)
-\frac{1}{2} \nabla_C \nabla_A \left(\nabla^C \phi \nabla_B \phi\right) \nonumber\\
- &\left.\frac{1}{2} \nabla_C \nabla_B \left(\nabla^C \phi \nabla_A \phi\right)
+ \frac{1}{2} \nabla_E \nabla^E \left( \nabla_A \phi \nabla_B \phi \right)\right],  
\end{align}
is the variation of the terms that depend on scalar field $\phi$ with respect to metric $g_{AB}$.

Variation of the action with respect to the scalar field gives us the scalar field equation of motion as
\begin{equation}
  g^{AB} \nabla_A \nabla_B \phi + \nabla_A \left[\nabla_B \phi \left(\xi g^{AB} R + \eta R^{AB}\right)\right] = 0.
\end{equation}

The metric for five dimensional universal extradimensions (5D UED) model is \cite{Bringmann:2003sz},
\begin{align}
ds^2 
&= - dt^2 + a^2 \delta_{ij}  dx^i dx^j + b^2 dz^2, 
\end{align}
where  $a=a(t)$ is the scale factor for  three dimensional (3D) common space  and $b=b(t)$ is the scale factor for the extra dimension. Here, we choose different scale factors $a$ and $b$ for the the 3D and extra dimension space for simplicity and we choose a flat metric for the 3D space to include recent observational results that our universe is flat \cite{spergel:2006hy}. 
 
From the metric, we get some geometrical quantities i.e the Ricci tensors
\begin{align}
R_{00} &= - \left(3\dot{H}_a + \dot{H}_b + 3H_a^2 + H_b^2 \right)\\
R_{ij} &= a^2 \delta_{ij} \left(\dot{H}_a +3H_a^2 + H_a H_b \right)\\
R_{55} &= b^2 \left(\dot{H}_b + 3H_aH_b + H_b^2\right),
\end{align}
Ricci scalar,
\begin{align}
R = 2 \left(3\dot{H}_a + \dot{H}_b + 6H_a^2 + 3H_a H_b +H_b^2\right),
\end{align}
and Einstein tensors 
\begin{align}
G_{00} &=  3H_a\left(H_a +H_b\right),\\
G_{ij} &=  -a^2 \delta_{ij}\left( 2\dot{H}_a +3H_a^2 + \dot{H}_b + H_b^2 + 2 H_a H_b\right),\\
G_{55} &=  -3b^2\left(\dot{H}_a + 2H_a^2 \right),
\end{align}
where $H_a \equiv \frac{\dot{a}}{a}$ and $H_b \equiv \frac{\dot{b}}{b}$ are Hubble parameter for three dimensional and extra spaces.
The overdot notation defined as derivative with respect to time $t$.

\section{Field Equation and Its Solution}
Now, we are ready to solve the Einsteins equations (\ref{eq:einstein}).
First of all, we make two assumptions concerning the scalar field $\phi$ and scale factors $a(t)$ and $b(t)$. First, we consider the scalar that depends on time only, $\phi=\phi(t)$. Second, we impose the condition that the two scale factors related  as $b=a^\gamma$ for $\gamma$ constant, as taken in \cite{Arianto2011}. This relation gives us $H_b = \gamma H_a \equiv \gamma H$.

Substituting  geometrical quantities in the last section into eq.(\ref{eq:einstein}) and applying the two assumptions, we get
\begin{align}
\label{eq:einstein00}
  3\left(1+\gamma\right) H^2 &= \left(\frac{1}{2} +\alpha_1 H^2 +\alpha_2 \dot{H} \right)
  \kappa^2\dot{\phi}^2 \nonumber\\
  &\qquad- \alpha_2 H \kappa^2 \dot{\phi}\ddot{\phi}, \\
\label{eq:einstein11}
  -\left(\beta_1 H^2 +\beta_2 \dot{H} \right) & = \left[\frac{1}{2} + \left(\xi +\eta\right)\left(\beta_1 H^2 +\beta_2 \dot{H}\right) \right]\kappa^2\dot{\phi}^2 \nonumber\\
  +&\beta_3 H \kappa^2\dot{\phi}\ddot{\phi} +\beta_4 \kappa^2\left(\dot{\phi}\dddot{\phi} + \ddot{\phi}^2 \right),\\
\label{eq:einstein55}
  -3\left(2H^2 +\dot{H}\right) &= \left[\frac{1}{2} +3\left(\xi+\eta\right)\left(2H^2 +\dot{H}\right) \right]\kappa^2\dot{\phi}^2 \nonumber\\
  +&6\left(\xi+\eta\right) H \kappa^2\dot{\phi}\ddot{\phi} 
  +\beta_4 \kappa^2\left(\dot{\phi}\dddot{\phi} + \ddot{\phi}^2 \right),
\end{align} 
where
\begin{align}
  \alpha_1 &\equiv \gamma^2 \left(2\xi +\eta\right) + 3\left(3\xi -\gamma \eta\right),\nonumber\\
  \alpha_2 &\equiv \left(3+\gamma\right)\left(2\xi+\eta\right), \nonumber\\
  \beta_1 &\equiv\left(\gamma^2 + 2\gamma +3\right),\\
  \beta_2 &\equiv\left(\gamma + 2 \right),\nonumber\\
  \beta_3 &\equiv 2\left(\xi+\eta\right)\left(2+\gamma\right),\nonumber\\
  \beta_4 &\equiv 2\xi +\eta,\nonumber
\end{align}
are constants. 
The scalar field equation of motion reads,
\begin{align}
\label{eq:gerak}
   \ddot{\phi} \left(1+\epsilon_1 H^2 +\epsilon_2 \dot{H}\right)
   + \dot{\phi} \left[\epsilon_3 H + \epsilon_4 H^3 + \epsilon_5 H\dot{H} +\epsilon_2 \ddot{H} \right]
 = 0,
\end{align}
where,
\begin{align}
 \epsilon_1 &\equiv \left(\gamma^2 + 3\right)\left(2\xi+\eta\right) + 6\xi\left(\gamma +1 \right),\nonumber\\
 \epsilon_2 &\equiv \left(\gamma + 3\right)\left(2\xi+\eta\right),\nonumber\\
\label{eq:epsilon}
 \epsilon_3 &\equiv \gamma + 3,\\
 \epsilon_4 &\equiv \left(\gamma + 3\right)\epsilon_1,\nonumber \\
 \epsilon_5 &\equiv 3\left(2\xi +\eta\right)\left(\gamma^2+2\gamma+5\right)+12\xi\left(\gamma^2 + 1\right) \nonumber
\end{align}
are constants. 
 
 We see that the field equations contain the third derivative of the scalar field and the second derivative of the Hubble parameter. We can simplify the field equations by choosing 
 a specific condition for $\xi$ and $\eta$ parameters as follows
 \begin{equation}
 2\xi+\eta=0.
 \end{equation} 
So we have $\alpha_2 = \beta_4 = \epsilon_2 = 0$. We can also combine eqs. (\ref{eq:einstein11}) and (\ref{eq:einstein55}) as
\begin{align}
\label{eq:einstein1155}  
  -\left(\lambda_1 H^2 +\lambda_2 \dot{H}\right) = \left[{1}  - \xi \left(\lambda_1 H^2 +\lambda_2 \dot{H}\right) \right]\kappa^2\dot{\phi}^2 +\lambda_3 H \kappa^2\dot{\phi}\ddot{\phi},
\end{align} 
 where
 \begin{align}
   \lambda_1 &\equiv \gamma^2+2\gamma+9,\nonumber\\
  \label{eq:lambda}
   \lambda_2 &\equiv \gamma + 5,\\
   \lambda_3 &\equiv -\xi \left(2\gamma +10\right).\nonumber
 \end{align}
 
 As a special case, we solve the field equation for $\xi=0$. For this case, the field equations reads  
 \begin{align}
 \label{eq:einstein00sec1}
   -\left(1+\gamma\right) H^2 &= \frac{\kappa^2}{2}\dot{\phi}^2,\\
 \label{eq:einstein1155sec1}
  -\left(\lambda_1 H^2 + \lambda_2 \dot{H}\right) &= {\kappa^2}\dot{\phi}^2,\\
 \label{eq:geraksec1}
   \left(\gamma + 3\right) H \dot{\phi} + \ddot{\phi} &= 0.
 \end{align}
From eqs. (\ref{eq:einstein00sec1}) and (\ref{eq:geraksec1}) we get
\begin{align}
\label{eq:phi1}
  {\phi} &= {\phi}_0 + \frac{\sqrt{-2\left(1+\gamma\right)}}{\kappa\left(3+\gamma\right)}\ln{\left(t-t_0\right)}, \\
  a &= a_0 \left(t-t_0\right)^{\frac{1}{3+\gamma}},
\end{align}
where $\dot{\phi}_0$, $a_0$, and $t_0$ are constants. Here, for a static extradimension ($\gamma =0$), we have
$ a = a_0 t^{1/3}$, which is the standard cosmological solution with stiff matter.

Now we are going to study a general case of $\xi \neq 0$. For this case, eq. (\ref{eq:einstein00}) gives 
\begin{equation}
H^2 = \frac{\kappa^2 \dot{\phi}^2}{6\left(1+\gamma\right)-2\alpha_1\kappa^2\dot{\phi}^2}.
\end{equation}
or, equivalently
\begin{equation}
\label{eq:phi0}
\dot{\phi}^2 = \frac{6\left(1+\gamma\right)H^2}{\kappa^2\left(1 +2\alpha_1 H^2\right)},
\end{equation}
Thus, we have two conditions for $H$ and $\dot{\phi}$ as follow
\begin{align}
3\left(1 +\gamma\right) - \alpha_1 \kappa^2 \dot{\phi}^2 &> 0,\\
\frac{1 + 2\alpha_1 H^2}{1+\gamma} &> 0.
\end{align}

Solving eqs. (\ref{eq:einstein1155}) and (\ref{eq:gerak}) for $\dot{H}$ and substituting the value of $\dot{\phi}$ from (\ref{eq:phi0}), we get
\begin{equation}
\label{eq:Hdot}
\dot{H} = \frac{\left\{6\left(1 +\gamma\right)\left[1 -\left(\xi\lambda_1+\lambda_3\epsilon_3\right)H^2 \right] + \lambda_1 \left(1+2\alpha_1 H^2\right)\right\}\left(1 + \epsilon_1 H^2\right) H^2}{-\lambda_2 \left\{1 + \left[2\alpha_1 - 6\left(1+\gamma\right)\xi\right] H^2 \right\} \left(1 + \epsilon_1 H^2\right) + 6 \left(1 +\gamma\right)\lambda_3\epsilon_5 H^4}.
\end{equation}
First, lets check the power law solution, $a\propto t^p$, for a constant $p$. The last equation gives us
\begin{equation}
  p=\frac{(\gamma^2+8\gamma+15)p^2 t^4+(18\gamma^3+165\gamma^2+420\gamma+297)\xi p^4t^2+(72\gamma^4+738\gamma^3+2358\gamma^2+2934\gamma+1242)\xi^2p^6)}{(\gamma+5)t^4+(12\gamma^2+78\gamma+90)\xi p^2t^2+(144\gamma^4+900\gamma^3+1152\gamma^2+1476\gamma+1080)\xi^2p^4}
\end{equation}
One can verify that there is no $\gamma$ value that make $p$ constant for every $t$. Thus there is no power law solution that valid for all $t$. 
The solution of eq. (\ref{eq:Hdot}) is the  de Sitter solution $e^{Ht}$ where 
\begin{equation}
\label{eq:deSitter}
H = \left(-\epsilon_1\right)^{-1/2}=\frac{1}{\sqrt{-6\xi\left(\gamma +1\right)}}.
\end{equation}
This result  also gives a constraint for $\xi$ and $\gamma$ i.e. $\xi\left(\gamma + 1\right) < 0$. 

Now, we are going to investigate two special cases of large and small Hubble parameter. 
For a large Hubble parameter, eq. (\ref{eq:Hdot}); after substituting the value of $\alpha_1$, $\epsilon$'s and $\lambda$'s; gives us
\begin{equation}
\dot{H} \approx - \frac{ \left(3\gamma^3+22\gamma^2+59\gamma+48\right)}{2\left(\gamma+1\right)\left(\gamma+5\right)\left[3\xi\left(\gamma+2\right)+\left(\gamma+3\right)\right]}\left[1+6\xi\left(\gamma+1\right) H^2\right],
\end{equation}
where the solution is 
\begin{equation}
\label{eq:Hbesar}
a=a_0 e^{\frac{t}{\varepsilon_1}} \left(e^{-\varepsilon_1 A t} -1\right)^{\frac{\varepsilon_1 A}{2}},
\end{equation}
with $A \equiv - \frac{ \left(3\gamma^3+22\gamma^2+59\gamma+48\right)}{2\left(\gamma+1\right)\left(\gamma+5\right)\left[3\xi\left(\gamma+2\right)+\left(\gamma+3\right)\right]}$.

Suppose that the Hubble parameter decreases. 
For $\xi>0$ and $\left[1+6\xi\left(\gamma+1\right) H^2\right]>0$, we found that $\dot{H}<0$ if 
$\gamma < -5$, $-3<\gamma<-2$ (with $\xi<-\frac{3\left(\gamma+2\right)}{\left(\gamma+3\right)}$), or $-2\leq\gamma<-1.42$.
Then, after the Hubble parameter reach a small value, eq. (\ref{eq:Hdot}) reduces to
\begin{equation}
\dot{H} = - \frac{\left[ \lambda_1 + 6\left(1+\gamma\right)\right] H^2}{\lambda_2 \left(1+2\alpha_1 H^2\right)} = - \frac{\left(\gamma +3\right)H^2}{1+2\alpha_1 H^2}.
\end{equation}
The solution for the scale factor is 
\begin{equation}
a = a_0 \left(t - t_0\right)^{\frac{1}{3+\gamma}},
\end{equation}
which is equivalent with the case of pure free scalar.

Finally, we are going to study a very special case of our theory where the scalar field is linear in $t$. For this case, we have $\kappa^2 \dot{\phi}^2 \equiv \psi_0$ constant. The field equations can be written as
\begin{align}
\label{eq:linear00}
 3\left(1+\gamma\right)H^2 &= \left(\frac{1}{2} +\alpha_1 H^2 \right)\psi_0,\\
 \label{eq:linear1155}
 -\left(\lambda_1 H^2 +\lambda_2 \dot{H}\right) &= \left[1 -\xi \left(\lambda_1 H^2 +\lambda \dot{H}\right)\right] \psi_0,\\
 \label{eq:lineargerak}
 \epsilon_3 +\epsilon_4 H^2 +\epsilon_5 \dot{H} &= 0.
\end{align}
From eq. (\ref{eq:linear00}) we get constant $H^2$, then eq. (\ref{eq:lineargerak}) gives
\begin{equation}
\label{eq:linearH}
  H^2 = -\frac{1}{6\xi\left(\gamma+1\right)},
\end{equation}
which is equivalent to eq. (\ref{eq:deSitter}).
Substituting this result to eq. (\ref{eq:linear00}), we have
\begin{equation}
\label{eq:linearPhi}
  \phi = \phi_0 + \frac{1}{\kappa} \sqrt{\frac{1 +\gamma}{\xi\left(2+\gamma\right)}} t.
\end{equation}
Requiring $\phi$ to be a real field, we have $\gamma<-2$ or $\gamma>-1$  for $\xi > 0$  and $-2<\gamma<-1$ for $\xi < 0$.

\section{Stability Analysis}
In this section, we study the stability  of dynamical system of this model. Introducing three new variables,
\begin{align}
  x &= \frac{\kappa \dot{\phi}}{H}, \\
  y &= {\kappa \dot{\phi}},\\
  N &= \ln a,
\end{align}
the field equations (\ref{eq:einstein00}) to (\ref{eq:einstein1155}) dan be written as
\begin{align}
1 &=  \frac{x^2}{6\left(1+\gamma\right)} + 3\xi y^2 ,\\
\label{eq:x}
y' &= -y \left(\frac{\varepsilon_5 x^2 y^2 + \left(1-\xi y^2\right)
\left[\left(\lambda_1 \varepsilon_5-\lambda_2\varepsilon_4\right)y^2-\lambda_2 \varepsilon_3 x^2\right]}{\lambda_3 \varepsilon_5 y^2 - \lambda_2 \left(1-\xi y^2\right)\left(x^2+\varepsilon_1 y^2\right)} \right),\\
\label{eq:y}
x' &= \frac{x}{\varepsilon_5 y^2} \left\{ - \left(\frac{\varepsilon_5 x^2 y^2 + \left(1-\xi y^2\right)
\left[\left(\lambda_1 \varepsilon_5-\lambda_2\varepsilon_4\right)y^2-\lambda_2 \varepsilon_3 x^2\right]}{\lambda_3 \varepsilon_5 y^2 - \lambda_2 \left(1-\xi y^2\right)\left(x^2+\varepsilon_1 y^2\right)} \right)\left[ x^2 + \left(\varepsilon_1 +\varepsilon_5\right)y^2\right] \right.\nonumber\\
&~~\left.+ \left(\varepsilon_3 x^2 + \varepsilon_4 y^2\right)\right\},
\end{align}
where the prime denotes the first derivative with respect to $N$.

The last two equation are an autonomous phase system of the form $\dot{\textbf{x}} = \textbf{f}(\textbf{x})$ where $\textbf{x}=(x,y)$. The critical points $\textbf{x}_0$ of the system  comes from the solutions of $\textbf{f}(\textbf{x}_0) = 0$. The stability of the critical points are determined by the eigenvalues $\mu_i$ of the matrix $\mathcal{M}$ whose components are $\mathcal{M}_{ij} \equiv \left.\frac{\partial f_i}{\partial x_j}\right|_{\textbf{x}_0}$. A critical point is called stable (unstable) if $Re\left(\mu_i\right) <0$ ($Re\left(\mu_i\right) >0$) for all $i$, otherwise it called a saddle point \cite{Copeland:2006wr}.

The first critical point of eqs. (\ref{eq:x}) and (\ref{eq:y}) is $\mathcal{P} = (0,0)$. The $\mathcal{M}$ matrix does not have any eigenvalue at $\mathcal{P}$, because the  function $\frac{\partial f_i}{\partial x_j}$ are discontinuous at $\mathcal{P}$. Thus, the point $\mathcal{P}$ is a saddle point.
The second critical point  is $\mathcal{Q} = (x_1, y_1)$, where
\begin{align}
  y_1^2 &= \frac{1}{\xi + \Gamma},\\
  x_1^2 &= -\frac{\varepsilon_4}{\varepsilon_3} y^2,
\end{align}
and $\Gamma \equiv \frac{\varepsilon_1 \varepsilon_5}{\lambda_1\varepsilon_5 - \lambda_2 \varepsilon_4 +\lambda_2 \varepsilon_1 \varepsilon_3}$. This point  corresponds to the de Sitter solution, $\dot{H} =0$, as in eq. (\ref{eq:Hdot}). Recalling the value of $\varepsilon$'s and $\lambda$'s from (\ref{eq:epsilon}) and (\ref{eq:lambda}), we have
\begin{align}
  y_1^2 &= \frac{\gamma^2 +2 \gamma +9}{\left(\gamma +3 \right)\left(\gamma +5\right)\xi},\\
  x_1^2 &= -\frac{6\left(\gamma+1\right)\left(\gamma^2 +2 \gamma +9\right)}{\left(\gamma +3 \right)\left(\gamma +5\right)}.
\end{align}
Hence, in order to have real $(x_1, y_1)$, we require
\begin{align}
\label{eq:xi}
  \xi &>0,\\
  \label{eq:gamma}
  \gamma &< -5 ~~\textrm{or}~~ -3<\gamma<-1.
\end{align}

Next, we will consider the stability of the $\mathcal{Q}$ point by choosing a particular value of $\gamma$. For example, we take $\gamma =-2$, and the component of $\mathcal{M}$ matrix are
\begin{equation}
\mathcal{M} = \left(
\begin{array}{cc}
f_x &f_y\\
g_x &g_y
\end{array}
 \right) = \left(
\begin{array}{cc}
-\frac{21.13}{\sqrt{\xi}} &\frac{0.48\left(10-\xi\right)}{\sqrt{\xi}}\\
-\frac{5.062}{\sqrt{\xi}} &\frac{2.60-9.2\sqrt{\xi}}{\xi\sqrt{\xi}}
\end{array}
 \right).
\end{equation}
The eigenvalues of the $\mathcal{M}$ are determined by the characteristic equation
\begin{equation}
\mathcal{A}_1 \mu^2 + \mathcal{B}_1 \mu + \mathcal{C}_1 = 0,
\end{equation}
where $\mathcal{A}_1=1$, $\mathcal{B}_1 = -\left(f_x + g_y\right)$ and $\mathcal{C}_1 = f_x g_y - f_y g_x$. 
The characteristic equation will give us two negative real roots if and only if $\mathcal{D}_1 = \mathcal{B}_1^2 - 4\mathcal{A}_1 \mathcal{C}_1 >0$, $\mathcal{B}_1 > 0$, and $\mathcal{C}_1 >0$. For $\gamma =-2$,  $\mathcal{D}_1$ and $\mathcal{B}_1$ are positive definite, and $\mathcal{C}_1$ will be positive for $1.108 <\xi<1.79$. 
For instant, if we choose $\xi=1.6$, we have $\mu_1 = -0.056$ and $\mu_2 = -15.94$. Hence the point  $\mathcal{Q}$  is stable.

\section{Conclusion}
We have studied the nonminimal derivative coupling of scalar field and curvature tensor in five dimensional universal extra dimension model. Here, we choose $2\xi +\eta=0$ condition to remove  higher order derivatives of the scalar field.  For a model with only free kinetic term of scalar field, we find the scalar field evolves with  velocity  proportional to $t^{-1}$ and the scale factor evolves as $a \propto \left(t-t_0\right)^{1/(3+\gamma)}$. Here, we see that a constant extra dimension ($\gamma = 0$)  will give us a solution of $a \propto \left(t-t_0\right)^{1/3}$ which is equal to the standard cosmological solution with stiff matter. 

In the general case of pure free scalar field with NMDC, the Hubble parameter evolves as given by (\ref{eq:Hdot}), from which we get the de Sitter solution with the value of cosmological constant  given by ${\Lambda} = {3}H^2 = - \frac{1}{2\xi\left(\gamma+1\right)}$.  This solution is equivalent to the case of linear scalar field. We can also choose the $\gamma$ and $\xi$ parameters, so that the Hubble parameter decreases.  In the limit of large Hubble parameter, the scale factor evolves as given in (\ref{eq:Hbesar}) while in the limit of a small Hubble parameter, the scale factor is equivalent to that of the pure free scalar case. Thus for a static extra dimension ($\gamma =0$) our results are exactly the same as the results of four dimensional case \cite{sushkov}. But, according to our stability analysis, the case of static extra dimension have no stable point. In order to have a real stable critical point, we require negative $\gamma$ as in eq. (\ref{eq:gamma}). 
It means that the extra dimension, if it exists, have to be shrunk by the expansion of the four dimensional spacetime. The stable critical point of our model  corresponds to de Sitter solution.


\section*{Acknowledgments}
This work is supported by Excellent Doctoral Fellowship of the Faculty of Mathematics and Natural Sciences Institut Teknologi Bandung (ITB), Riset Peningkatan Kapasitas ITB 2012, Riset dan Inovasi KK ITB 2012-2013, and Hibah Desentralisasi Dikti 2012. This paper is tributed to Arianto who passed away on July 14, 2011. We acknowledged him for useful discussion in early stage of this work. We  thank Jusak S. Kosasih for careful reading and correcting
English grammar.


\begin{thebibliography}{10}

\bibitem{Overduin:1998pn}
J.~M. Overduin and P.~S. Wesson.
\newblock {Kaluza-Klein gravity}.
\newblock {\em Phys. Rept.}, 283:303--380, 1997.

\bibitem{maartens}
Roy Maartens and Kazuya Koyama.
\newblock {Brane-World Gravity}.
\newblock {\em Living Rev. Rel.}, 13:5, 2010.

\bibitem{RSI}
Lisa Randall and Raman Sundrum.
\newblock {A large mass hierarchy from a small extra dimension}.
\newblock {\em Phys. Rev. Lett.}, 83:3370--3373, 1999.

\bibitem{RSII}
Lisa Randall and Raman Sundrum.
\newblock {An alternative to compactification}.
\newblock {\em Phys. Rev. Lett.}, 83:4690--4693, 1999.

\bibitem{SMS}
Tetsuya Shiromizu, Kei-ichi Maeda, and Misao Sasaki.
\newblock {The Einstein equations on the 3-brane world}.
\newblock {\em Phys. Rev.}, D62:024012, 2000.

\bibitem{Brax:2003fv}
Philippe Brax and Carsten van~de Bruck.
\newblock {Cosmology and brane worlds: A review}.
\newblock {\em Class. Quant. Grav.}, 20:R201--R232, 2003.

\bibitem{Langlois:2002bb}
David Langlois.
\newblock {Brane cosmology: An introduction}.
\newblock {\em Prog. Theor. Phys. Suppl.}, 148:181--212, 2003.

\bibitem{arianto1}
Arianto, Freddy~P. Zen, and Bobby~Eka Gunara.
\newblock {Modified Gravitational Equations on Braneworld with Lorentz
  Invariant Violation}.
\newblock {\em Gen. Rel. Grav.}, 42:909--927, 2010.

\bibitem{arianto}
Arianto, Freddy~P. Zen, Bobby~Eka Gunara, Triyanta, and Supardi.
\newblock {Some Impacts of Lorentz Violation on Cosmology}.
\newblock {\em JHEP}, 09:048, 2007.

\bibitem{arianto2}
Arianto, Freddy~P. Zen, Triyanta, and Bobby~Eka Gunara.
\newblock {Attractor Solutions in Lorentz Violating Scalar-Vector- Tensor
  Theory}.
\newblock {\em Phys. Rev.}, D77:123517, 2008.

\bibitem{zen}
Freddy~P. Zen, Arianto, Bobby~Eka Gunara, Triyanta, and A.~Purwanto.
\newblock {Cosmological evolution of interacting dark energy in Lorentz
  violation}.
\newblock {\em Eur. Phys. J.}, C63:477--490, 2009.

\bibitem{UED}
Thomas Appelquist, Hsin-Chia Cheng, and Bogdan~A. Dobrescu.
\newblock {Bounds on universal extra dimensions}.
\newblock {\em Phys. Rev.}, D64:035002, 2001.

\bibitem{Hsieh:2006qe}
Ken Hsieh, R.N. Mohapatra, and Salah Nasri.
\newblock {Dark matter in universal extra dimension models: Kaluza-Klein photon
  and right-handed neutrino admixture}.
\newblock {\em Phys.Rev.}, D74:066004, 2006.

\bibitem{Appelquist:2001mj}
Thomas Appelquist, Bogdan~A. Dobrescu, Eduardo Ponton, and Ho-Ung Yee.
\newblock {Proton stability in six-dimensions}.
\newblock {\em Phys.Rev.Lett.}, 87:181802, 2001.

\bibitem{Bringmann:2003sz}
Torsten Bringmann, Martin Eriksson, and Michael Gustafsson.
\newblock {Cosmological evolution of universal extra dimensions}.
\newblock {\em Phys. Rev.}, D68:063516, 2003.

\bibitem{Bringmann:2004xt}
Torsten Bringmann, Martin Eriksson, and Michael Gustafsson.
\newblock {Stability of homogeneous extra dimensions}.
\newblock {\em AIP Conf.Proc.}, 736:141--146, 2005.

\bibitem{amendola}
Luca Amendola.
\newblock {Cosmology with nonminimal derivative couplings}.
\newblock {\em Phys. Lett.}, B301:175--182, 1993.

\bibitem{Capozziello:1999xt}
S.~Capozziello, G.~Lambiase, and H.~J. Schmidt.
\newblock {Nonminimal derivative couplings and inflation in generalized
  theories of gravity}.
\newblock {\em Annalen Phys.}, 9:39--48, 2000.

\bibitem{capozziello2}
S.~Capozziello and G.~Lambiase.
\newblock {Nonminimal derivative coupling and the recovering of cosmological
  constant}.
\newblock {\em Gen. Rel. Grav.}, 31:1005--1014, 1999.

\bibitem{granda}
L.~N. Granda.
\newblock {Non-minimal Kinetic coupling to gravity and accelerated expansion}.
\newblock {\em JCAP}, 1007:006, 2010.

\bibitem{sushkov}
Sergey~V. Sushkov.
\newblock {Exact cosmological solutions with nonminimal derivative coupling}.
\newblock {\em Phys. Rev.}, D80:103505, 2009.

\bibitem{spergel:2006hy}
D.~N. Spergel et~al.
\newblock {Wilkinson Microwave Anisotropy Probe (WMAP) three year results:
  Implications for cosmology}.
\newblock {\em Astrophys. J. Suppl.}, 170:377, 2007.

\bibitem{Arianto2011}
Arianto, F.P. Zen, S.~Feranie, I~P. Widyatmika, and B.E. Gunara.
\newblock {Kaluza-Klein brane cosmology with a bulk scalar field}.
\newblock {\em Phys.Rev.}, D84:044008, 2011.

\bibitem{Copeland:2006wr}
Edmund~J. Copeland, M.~Sami, and Shinji Tsujikawa.
\newblock {Dynamics of dark energy}.
\newblock {\em Int.J.Mod.Phys.}, D15:1753--1936, 2006.

\end{thebibliography}
\end{document}